\documentclass[12pt]{iopart}

\usepackage{graphicx}
\usepackage{psfrag}
\usepackage{psfig}
\newcommand{\nn}{\nonumber}
\newcommand{\be}[0]{\begin{eqnarray}}
\newcommand{\ee}[0]{\end{eqnarray}}

\begin{document}

\title{Boundary field induced first-order transition in the
2D Ising model: numerical study}

\author{Elmar Bittner and Wolfhard Janke}

\address{Institut f\"ur Theoretische Physik 
and  Centre for Theoretical Sciences (NTZ), 
Universit\"at Leipzig,
Postfach 100\,920, D-04009 Leipzig, Germany}
\ead{elmar.bittner@itp.uni-leipzig.de}
\ead{wolfhard.janke@itp.uni-leipzig.de}

\begin{abstract}
\noindent
In a recent paper, Clusel and Fortin [{\it J. Phys. A.: Math. Gen. {\bf 39} (2006) 995}] 
presented an analytical study of a first-order transition induced by an
inhomogeneous boundary magnetic field in the two-dimensional Ising model. They identified the
transition that separates the regime where the interface is localized near the
boundary from the one where it is propagating inside the bulk.
Inspired by these results, we measured the interface tension by 
using multimagnetic simulations combined with parallel tempering 
to determine the phase transition and the location of the interface. 
Our results are in very good agreement with the theoretical predictions.
Furthermore, we studied the spin-spin correlation function for which no
analytical results are available.
\end{abstract}

\maketitle
\section{Introduction} \label{intro}
Wetting transitions are phase transitions in the surface layer of bulk
systems which are induced by symmetry-breaking surface fields~\cite{cahn,ebner}.
The Ising model with a boundary magnetic field is a simple model for such a
wetting problem, because Ising ferromagnets have the same critical behaviour
as the analogous case of gas-fluid transitions, as has been pointed out by
Nakanishi and Fisher~\cite{nafi}.
The use of the Ising model with short range interactions for wetting studies
has not only the advantage that one can use all the advanced simulation techniques
which have been developed in the past years.
Especially in two dimensions (2D),
there are also a lot of theoretical results available for comparison. 

The Ising model with a uniform boundary magnetic field 
on one side of a square lattice has been completely solved by McCoy and Wu~\cite{mccoy},
whereas the Ising model with a uniform bulk field can only be solved 
at the critical temperature~\cite{zamo}. 
For situations with fixed boundary spins or 
equivalently infinite boundary magnetic fields~\cite{abra},
or finite boundary magnetic fields~\cite{auyang} some exact results have also been found.
In a recent paper, Clusel and Fortin~\cite{clusel1} presented an alternative method to
that developed by  McCoy and Wu for obtaining some exact results for the
2D Ising model with a general boundary magnetic field and for finite-size systems.
Their method is based on the fermion representation of the Ising model using a Grassmann 
algebra. They applied this method to study the first-order transition induced by an
inhomogeneous boundary magnetic field in the 2D Ising model~\cite{clufor}. 
To be more precise, the boundary magnetic field acts on the $x=1$ column of spins, 
being positive in the lower and negative in the upper halve.
By taking the thermodynamic limit exactly for a given geometry of the lattice,
they obtained a simple equation for the transition line and also a 
threshold for the aspect ratio $\zeta=L_x/L_y = 1/4$,
where this line moves into the complex plane. This vanishing of the transition line
indicates the crossover from 1D behaviour for $L_x\ll L_y$ to 2D behaviour at large $\zeta$,
which is reflected in the behaviour of the boundary spin-spin correlation function.  

The aim of this work is to check some of the predictions by carrying out
Monte Carlo simulations of this model and to extend the results to
parameter ranges and for observables where analytic solutions cannot be obtained. 
The rest of the paper is organized as follows. In \Sref{model} we give the 
definition of the model and briefly summarize the theoretical predictions.
A description of the employed simulation techniques and the results of our
Monte Carlo simulations are presented in \Sref{results}, 
and concluding remarks can be found in \Sref{summary}.

\section{Model and Theoretical Predictions} \label{model}
We consider a 2D Ising model with a non-homogeneous magnetic field
$h_y$ located on one boundary of the system. The Hamiltonian is given by
\begin{equation} 
{\cal H}=-J \sum_{x,y=1}^{L_x,L_y}(   \sigma_{x y}\sigma_{x+1 y} 
                            +  \sigma_{x y} \sigma_{x y+1})
      -\sum_{y=1}^{L_y}h_y\sigma_{1y}~,
\end{equation} 
with free boundaries in the $x$-direction and periodic boundary conditions
in the $y$-direction. To compare our results with the theoretical predictions
of Clusel and Fortin~\cite{clufor}, we consider the same profile of 
the boundary magnetic field acting on the $x=1$ column of spins: 
$h_y=H$ for $y=1,\dots,L_y/2$ and $h_y=-H$ for 
$y=L_y/2+1,\dots,L_y$, with $H\ge0$. 

\begin{figure}
\centerline{
\includegraphics[scale=0.85]{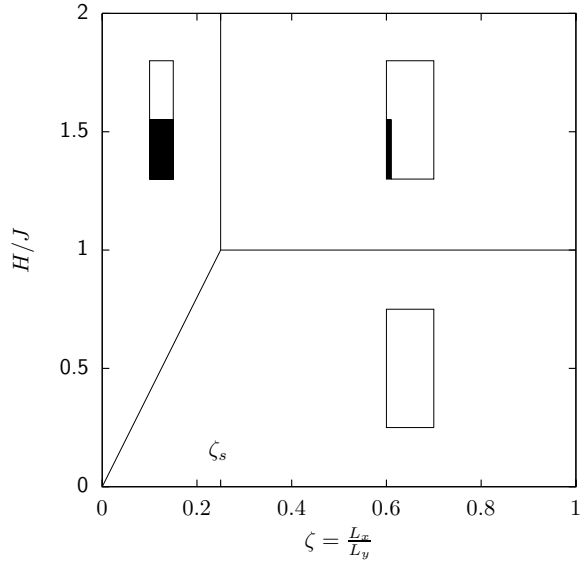}
}
\caption{The phase diagram at zero temperature as a function of the aspect ratio 
$\zeta=L_x/L_y$ and the boundary magnetic field  $H$.  
}
\label{fig_pd_t0}
\end{figure}

\begin{figure}
\centerline{
\includegraphics[scale=0.85]{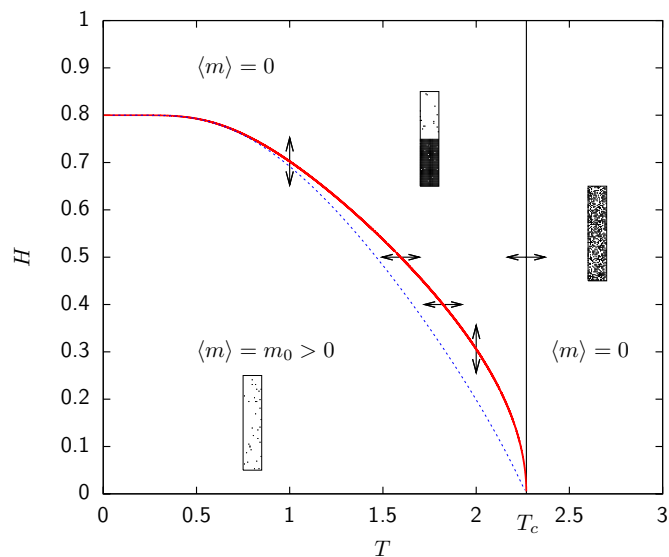}
}
\caption{The phase diagram for a system with $\zeta=0.2$. The thick line
shows the first-order transition given by Clusel and Fortin~\cite{clufor}
and the thin vertical line indicates the second-order bulk phase transition.
The thin dashed line indicates a simple approximation of the first-order line,
$H\approx2 \frac{\sigma(T) \zeta}{m_0(T)}$,
where $\sigma(T)$ and $m_0(T)$ are the interface tension and the spontaneous
magnetization of the pure 2D Ising model, respectively.
The sketches of the spin configurations illustrade the interface location in the 
three phases and the double-headed arrows show the parameters of the Monte Carlo simulations.
}
\label{fig_pd_zeta_0.2}
\end{figure}

In the limit of zero temperature, by using simple energetic arguments,
Clusel and Fortin~\cite{clufor} showed that for small $H$
all spins are aligned in one direction as in the bulk case for $H=0$, see also \Fref{fig_pd_t0}. 
With increasing $H$, however, depending on the aspect ratio
$\zeta=L_x/L_y$ two different interfaces can be formed. 
If $\zeta>\zeta_s=1/4$ the interface is localized near the boundary,
whereas for $\zeta<\zeta_s$ the interface is propagating inside the bulk.
The critical ratio $\zeta_s$ marks the
crossover from a 1D behaviour for $L_x\ll L_y$ towards a 2D behaviour
at large $\zeta$. 
For $\zeta<\zeta_s$ and non-zero temperatures $T>0$,  with the abbreviations
$t=\tanh(J/k_B T)$ and $u=\tanh(H/k_BT)$,
the equation for the first-order transition line in the $(t,u)$-plane
turns out to be a quadratic equation in $u^2$ ~\cite{clufor}:
\be \label{eq_2} \fl
&& 2t\Big(1+v(4\zeta)\Big)u^4+(1+t^2)\Big(1-2tv(4\zeta)-t^2\Big)u^2+2\Big(v(4\zeta)
-1\Big)t^3=0, \\ \nn
&& v(4\zeta)=\cosh \left [ 4\zeta\ln \left (\frac{1-t}{t(1+t)}\right ) \right ].
\ee
In \Fref{fig_pd_zeta_0.2} we
show the phase diagram for a system with aspect ratio $\zeta=0.2$ (and $J=k_B=1$).
In the low-temperature regime we can approximate the above expression
by comparing the energy of the interface with the energy induced by the
magnetic field. This leads to $H\approx 2 \sigma(T) \zeta /m_0(T)$, 
where $\sigma(T)$ and $m_0(T)$ are the known interface tension and the spontaneous
magnetization of the pure 2D Ising model, respectively.
This approximation 
reproduces the exact low-$T$ expansion, $H = 4\zeta - 4 \zeta T e^{-2/T}$, and
works very well for $T<1$ as one can see in \Fref{fig_pd_zeta_0.2}
(thin dashed line). Since $\sigma(T)$ vanishes much faster than $m_0(T)$ as
$T \rightarrow T_c$, also this point is reproduced exactly, but the slope
of the approximate transition line at $T_c$ does not diverge as for the
exact solution.

Due to the first-order transition induced by the inhomogeneous boundary 
magnetic field, the second-order phase transitions across the vertical 
line at $T=T_c$ are transitions from a region where an interface in the 
bulk separates two ordered domains of opposite magnetization from a 
disordered regime above the transition temperature. Therefore, the 
system undergoes a transition without a change in the magnetization 
$\langle m\rangle$ which is zero in both phases, cf. \Fref{fig_pd_zeta_0.2},
but the width of the magnetization distribution does change. 

\section{Numerical Results} \label{results}

Since we are primarily interested in the location of the interface induced 
by the boundary field, we first performed simulations at low temperatures 
to generate a well-defined interface. To overcome the slow dynamics at low
temperatures we developed a combination of the multimagnetic algorithm with
the parallel tempering method~\cite{wj_nic} for which we used two different
schemes: In the first scheme, we kept the magnetic field value $H$ fixed
and simulated $n=32$ replica of the system at different temperatures $T_i$.
In the second scheme, we kept the temperature $T$ fixed and used $n=32$ 
different values of the magnetic field $H_i$.

To construct the weight function for the multimagnetic part of the algorithm,
we employed an accumulative recursion, described in detail in Refs.~\cite{wj_nic}
and ~\cite{berg96}. Statistical averages were taken over runs of
$1\times 10^6$ Monte Carlo (MC) steps, where one MC step 
consists of one full multimagnetical lattice sweeps for all 32 replica and
one attempted parallel tempering exchange of all adjacent replica. With this
method we were able to study systems with $N=L_x\times L_y=50$ to $5000$ spins for 
aspect ratios $\zeta = L_x/L_y = 0.2$, $0.25 = \zeta_s$ and $0.5$, for further 
details see~\Tref{statistic}.

\begin{table*}[b]
  \caption{\label{statistic}Summary of simulation parameters (PT: parallel-tempering algorithm, SC: single-cluster update).
   }
 \begin{center}
  \begin{tabular}{cccccc}\hline\hline
   \makebox[1cm][c]{$\zeta$} &\makebox[2cm][c]{$H$} &\makebox[2cm][c]{$T$} &\makebox[2.5cm][c]{$L_x \times L_y$}&\makebox[2cm][c]{method} & \makebox[2.5cm][c]{measurements}\\ \hline
    0.2 & 0.4 & 1.6 -- 2.2 & 80 -- 2000 & PT& $1\times 10^6$\\
    0.2 & 0.5 & 1.4 -- 1.9 & 80 -- 2000 & PT&$1\times 10^6$\\
    0.2 & 0.5 & 2.1 -- 2.3 & 80 -- 180500 & SC&$1\times 10^6$ -- $5\times 10^6$\\
    0.2 & 0.7 -- 0.9 & 1.0 & 80 -- 640 &PT& $1\times 10^6$\\
    0.2 & 0.285 -- 0.316 & 2.0 & 80 -- 2000 &PT& $1\times 10^6$\\
    0.25 & 0.4 & 1.9 -- 2.1 &  256 -- 2500 &PT& $1\times 10^6$\\
    0.25 & 0.5 & 1.5 -- 1.9 &  64 -- 2000 & PT&$1\times 10^6$ \\
    0.25 & 0.5 & 2.0 -- 2.3 &  64 -- 6400 & SC&$1\times 10^6$ -- $5\times 10^6$ \\
    0.25 & 0.7 -- 0.8 & 1.5 &  64 -- 1600 & PT&$1\times 10^6$ \\
    0.5 & 0.5 & 2.2 -- 2.35 &  50 -- 3200 & PT&$1\times 10^6$ \\
    0.5 & 0.9 -- 1.1 & 1.0 &  50 -- 5000 & PT&$1\times 10^6$ \\
    0.5 & 1.5 -- 2.0 & 1.0 &  50 -- 3872 & PT&$1\times 10^6$ \\\hline\hline
  \end{tabular}
 \end{center}
\end{table*}

Let us first discuss the data obtained for the case $\zeta=0.2<\zeta_s$. 
For this value of the aspect ratio, the phase diagram as predicted 
by Clusel and Fortin~\cite{clufor} is shown in \Fref{fig_pd_zeta_0.2}. The
thick line indicates the first-order transitions from the fully magnetized
state with $\langle m \rangle >0$ to the mixed state with an interface 
extending across the bulk. To check the nature of these transitions we 
measured the probability density of the magnetization at four points along 
the transition line. In the first two cases, we kept the boundary magnetic 
field constant ($H=0.4$ and $0.5$) and varied the temperature to locate
the transition point, and in the other two cases, we fixed the temperature
($T=1.0$ and $2.0$) and varied the boundary magnetic field. These points are 
indicated by the double-headed arrows in \Fref{fig_pd_zeta_0.2}.

\begin{figure}
\centerline{
\includegraphics[scale=0.9]{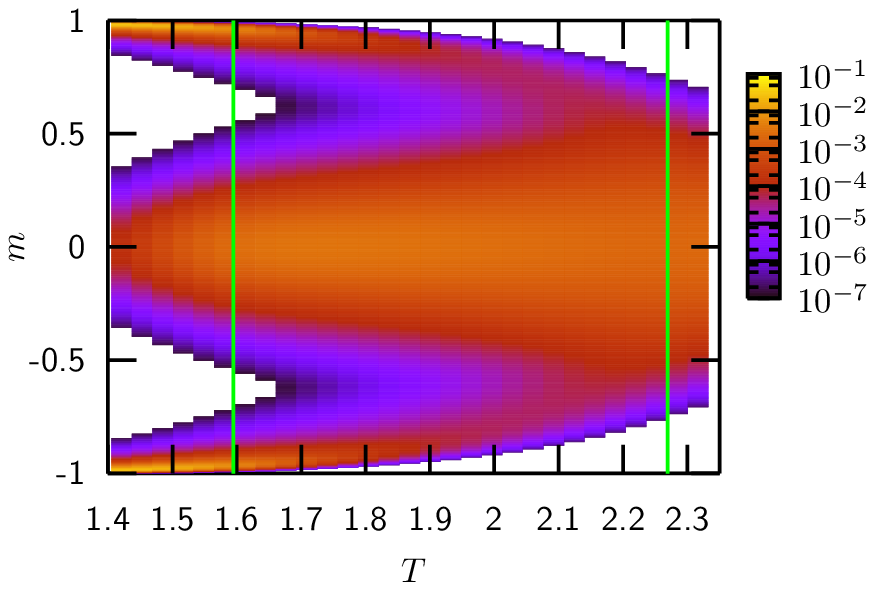}
\hspace{1.5cm}
\includegraphics[scale=1]{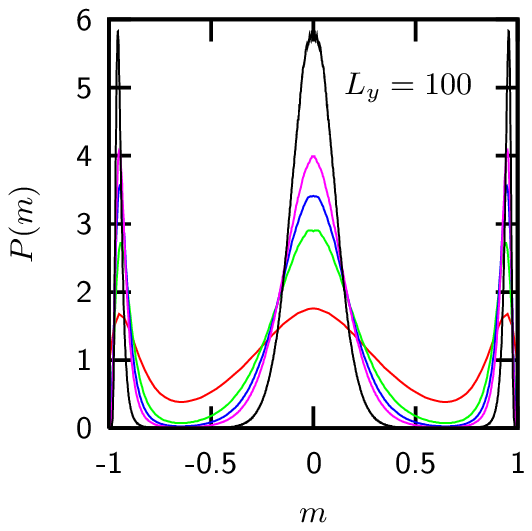}
}
\caption{{\em Left plot}: The probability density for the magnetization as a
function of temperature at constant boundary magnetic field $H=0.5$
for $N=20\times100$. The vertical lines indicate the first-order transition
temperature $T_0 \approx 1.5950$ according to \Eref{eq_2}
and the critical temperature $T_c\approx2.2692$ of the bulk phase transition.
{\em Right plot}: Histograms of the magnetization in the vicinity of $T_0$
for various lattice sizes ranging from $L_y=20$ to $L_y=100$.
Here the temperatures $T_0(L)$ are determined such that the peaks at
$m \approx \pm m_0$ and $m=0$ are of equal height.
}
\label{fig_3d_t2_z0.2}
\end{figure}

In the following we illustrate our procedure for obtaining the first-order
transition point and the associated interface tension for the case of
fixed $H=0.5$. A level plot of the magnetization density 
$m=(1/N)\sum_{x,y=1}^{L_x,L_y}\sigma_{x y}$ as a
function of temperature is shown in \Fref{fig_3d_t2_z0.2} (left). For each 
lattice size, a pseudo-transition point can be defined by varying the 
temperature until the peaks at $m \approx \pm m_0$ and $m=0$ are of equal height,
which can be achieved by histogram reweighting. The interface tension can then 
be estimated from \cite{1st_order}
\begin{equation}
F_L^s=\frac{1}{2 L}\ln{\left( \frac{P_L^{\rm max}}{P_L^{\rm min}}\right)},
\end{equation}
where $P_L^{\rm max}$ is the value of the peaks and $P_L^{\rm min}$ denotes 
the minimum in between, see \Fref{fig_3d_t2_z0.2} (right). The length of the
interface is denoted by $L$, which is $L=L_x$ in the case of $\zeta<\zeta_s$.  

The thus defined pseudo-transition temperatures $T_0(L)$ approach the 
infinite-volume transition temperature $T_0$ as $1/L^2$, 
and for the final estimate of $F^s=\lim_{L\rightarrow\infty}F_L^s$,
we performed a fit according to
\begin{equation}
F_L^s = F^s+\frac{a}{L}+\frac{b \ln(L)}{L}.
\label{eq:inter_fss}
\end{equation}
At fixed $T$ one proceeds analogously by varying the magnetic field $H$,
i.e., the roles of $T$ and $H$ are just interchanged.
For all four cuts at constant surface field or temperature we find a 
good agreement with the infinite-volume transition points derived 
from \Eref{eq_2} and a clearly nonzero interface tension, see \Tref{tab1}.

\begin{table}
  \caption{Results of the simulations close to the first-order transition
  line. For a given aspect ratio $\zeta$ we kept either the boundary magnetic 
  field $H$ or the temperature $T$ fixed. The third column shows the 
  measured transition points and the fourth column the exact 
  infinite-volume values given by \Eref{eq_2}. The fifth column contains
  our numerical estimates for the extrapolated interface tensions.
  }
  \label{tab1}
  \begin{center}
    \begin{tabular}{ccccc}
      \br 
      $\zeta$   & $H$ & $T_0$ & $T_0$ (exact) & $F_s$ \\ \hline
      0.2  & 0.4 & 1.84(1) & 1.82252$\dots$  &0.18(1)\\
      0.2  & 0.5 & 1.60(1) & 1.59497$\dots$  &0.32(2)\\
      0.25 & 0.5 & 1.91(1) & 1.95845$\dots$  &0.09(1)\\ \hline
     $\zeta$   & $T$ & $H_0$ & $H_0$ (exact) & $F_s$ \\ \hline
      0.2  & 1.0 & 0.72(1) &  0.702352$\dots$  &0.82(2)\\ 
      0.2  & 2.0 & 0.305(3) & 0.305928$\dots$  &0.12(1)\\
      0.25 & 1.5 & 0.73(1) &  0.762807$\dots$  &0.24(1)\\
      \br 
    \end{tabular}
  \end{center}
\end{table}

We also checked the critical behaviour along the line of second-order
transitions at $T=T_c=2/\log(1+\sqrt{2})\approx 2.2692$. To this end we 
run at $H=0.5$ single-cluster simulations (suitably adapted to the surface field)
for systems with $N=L_x\times L_y=4 \times 20$ to $190 \times 950$ spins and
performed a finite-size scaling (FSS) analysis to determine the transition point
and some critical exponents. This particular value of the magnetic field
has been chosen because of the relatively large temperature gap between
the boundary field induced first-order transition and the bulk phase 
transition. Between each measurement we performed one sweep, which here
consists of $n$ single-cluster updates with $n$ chosen such that 
$n \langle |{\cal S}| \rangle \approx N$, where $\langle |{\cal S}| \rangle$
is the average cluster size. For every run we generated $10^6$ sweeps, 
and recorded the time series of the energy density $e=E/N$ and
the magnetization density.
Using these time series, we can compute the specific heat,
$C=N (\langle e^2 \rangle- \langle e\rangle^2)/T^2$,
the (finite lattice) susceptibility,
$\chi=N (\langle m^2 \rangle-\langle m \rangle^2)$,
and the Binder cumulant $U= 1-\langle m^4 \rangle/3\langle m^2 \rangle^2$
in the vicinity of the simulation point by reweighting.

\begin{figure}
\centerline{ 
\includegraphics{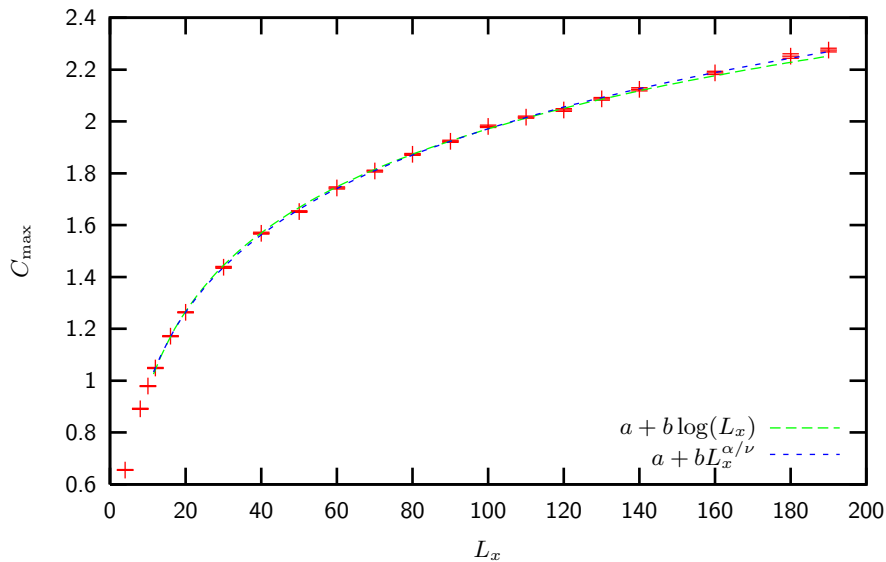}
}
\caption{
FSS of the specific-heat maxima $C_{\rm max}$. The logarithmic fit $C_{\rm max}=a+b \log(L_x)$ 
and an unbiased fit using the power-law ansatz $C_{\rm max}=a+b L_x^{\alpha/\nu}$ 
are almost indistinguishable on the scale of the figure.
}
\label{specificheat}
\end{figure}

In this way we can use the maxima of the (finite lattice) susceptibility 
to detect the pseudo-critical points and can obtain an estimate for 
$T_c$ from a linear least-square fit of their scaling behaviour,
$T_{\rm max} - T_c \propto L_x^{-1/\nu}=L_x^{-1}$, assuming thus the exact 
value $\nu=1$ according to the universality class of the 2D Ising model.
This leads to an estimate for the critical temperature,
$T_c = 2.2695(7)$,
which is in very good agreement with the exact value. The FSS ansatz 
for the (finite lattice) susceptibility maxima $\chi_{\rm max}$ is taken 
as usual as $\chi_{\rm max} \propto L_x^{\gamma/\nu}$. From a (linear) 
least-square fit, we find that $\gamma/\nu=1.75(4)$ is in perfect 
agreement with the exact value $7/4$. Concerning the specific heat we 
expect in the case of the Onsager exponent $\alpha=0$ a logarithmic
divergence of the form $C_{\rm max}=a+b \log(L_x)$. Indeed, the data 
can be fitted nicely with this ansatz, cf.~\Fref{specificheat}.  We also tried an unbiased fit
using the power-law ansatz $C_{\rm max}=a+b L_x^{\alpha/\nu}$, which 
gives us $\alpha=0.05(2)$, verifying the expected value.

\begin{figure}[t]
\centerline{ 
\includegraphics[scale=1.15]{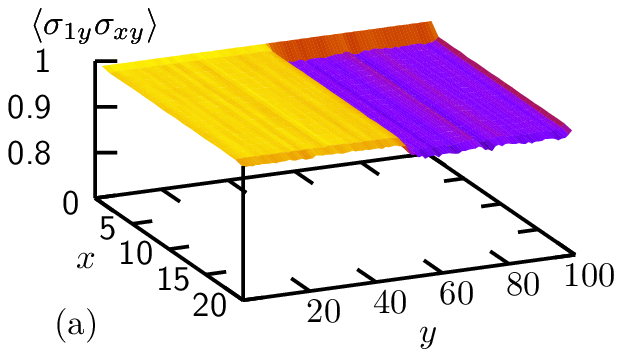}
\includegraphics[scale=1.15]{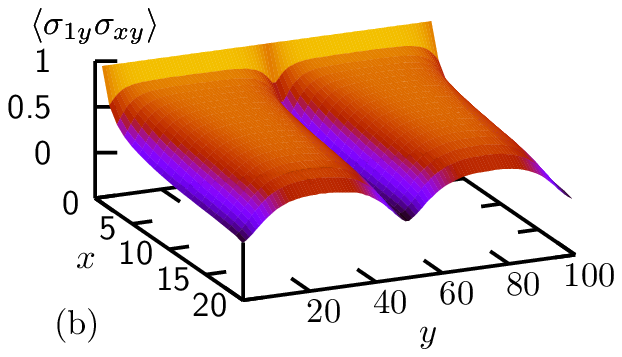}
}
\centerline{
\includegraphics[scale=1.15]{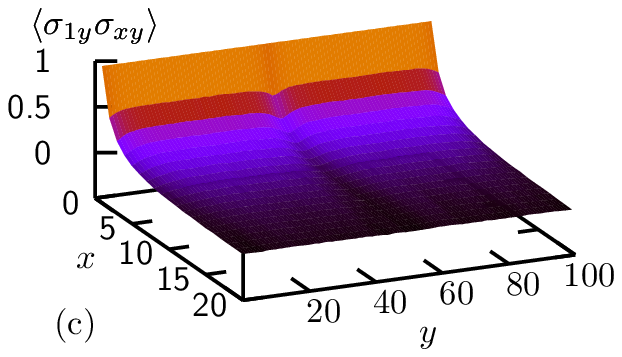} 
}
\caption{Spin-spin correlation function $\langle \sigma_{1y}\sigma_{xy} \rangle$ for $\zeta=0.2$, $L_y=100$, and
$H=0.5$ measured in different regimes:
(a) ordered regime at $T=1.0$,
(b) regime with an interface in the bulk at $T=2.0$, and
(c) disordered regime at $T=2.5$.  
 
}
\label{fig_spin_spin}
\end{figure}

For the aspect ratio $\zeta=0.2$, we also sampled the spin-spin correlation
functions $\langle \sigma_{1y} \sigma_{xy}\rangle$ and 
$\langle \sigma_{x1} \sigma_{xy}\rangle$ in $x$- and $y$-direction, respectively.
In the ordered regime where nearly all spins are aligned in the same direction ($\sigma_{xy}= +1$, say)
we find for $\langle \sigma_{1y} \sigma_{xy}\rangle$ and $1\le y \le L_y/2$ (where $h_y=H>0$)
almost constant values near unity as one expects. The fast decay of the spin-spin correlation
function to a slightly smaller value in the upper half of the system indicates that 
the interface is localized near the boundary, cf. \Fref{fig_spin_spin} (a).
In the regime with an interface in the bulk along the $x$-direction we find a symmetric shape of
$\langle \sigma_{1y} \sigma_{xy}\rangle$ as a function of $y$ which is a clear signal for the phase separation,
see \Fref{fig_spin_spin} (b). The opening angle between the plus and minus phases
for $1\le y\le L_y/2$ and $L_y/2+1 \le y \le L_y$, respectively, is a measure
for the fluctuations of the interface, e.g. a stiff interface shows an acute angle. 
For temperatures above the critical temperature, i.e. in the disordered phase,
we observe a similar vanishing of the spin-spin correlation functions as 
in the pure 2D Ising model, see \Fref{fig_spin_spin} (c).

\begin{figure}[t]
\centerline{
\includegraphics[scale=0.75]{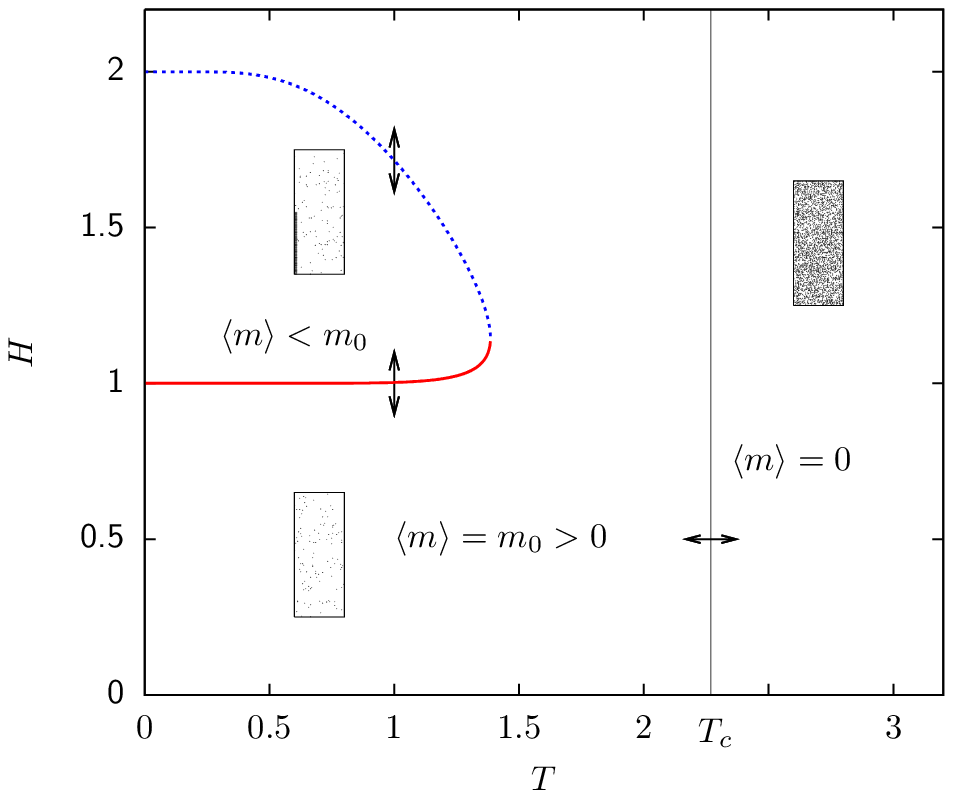}
\hspace{1.0cm}
\includegraphics[scale=0.75]{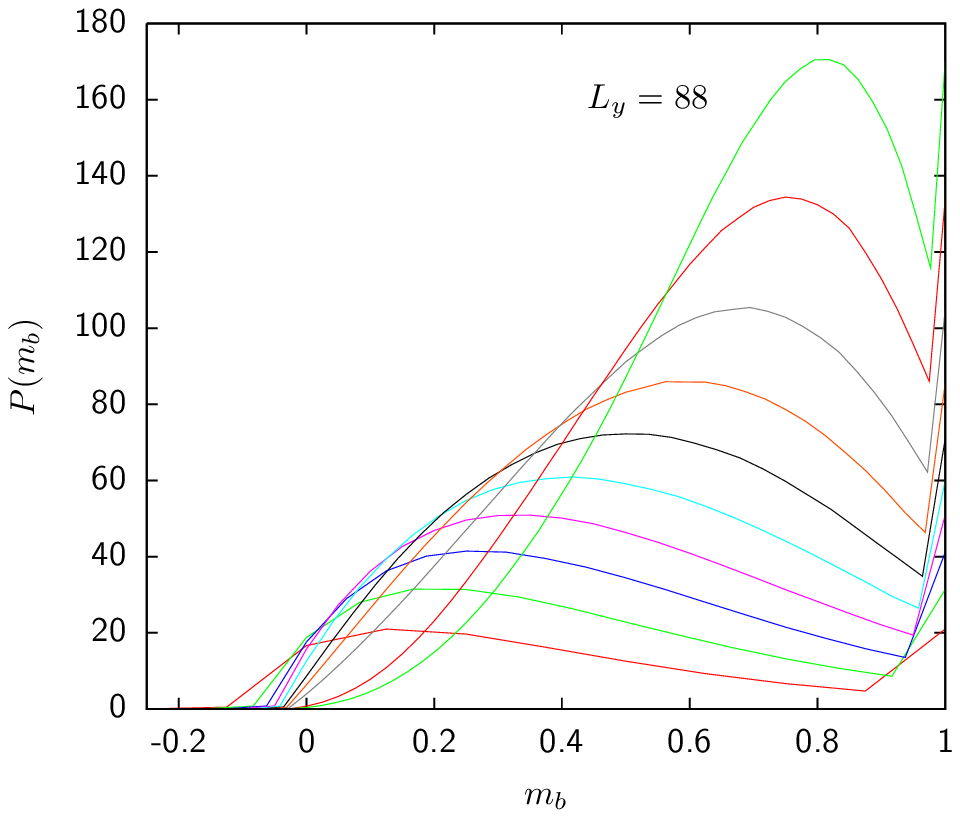}
}
\caption{{\em Left plot}: The finite-size phase diagram for a system with 
$\zeta=0.5$. The thick line shows the transition line between the ordered 
phase and the region with an interface localized on the boundary. The 
dashed line indicates  where a configuration with no interface  
and a configuration with an interface propagating inside the bulk have the same free energy 
(non-physical solution of \Eref{eq_2}, because this bulk interface configuration has a 
higher free energy then the one with an interface localized on the boundary and hence is 
suppressed.).
The thin vertical line indicates the second-order bulk phase transition.
{\em Right plot}: Probability density of the boundary magnetization $m_b$ for $T=1.0$ plotted
for various lattice sizes ranging from $L_y=16$ to $L_y=88$ 
and boundary magnetic field values $H \approx 1$ 
where the peaks are of equal height.
}
\label{fig_pd_zeta_0.5}
\end{figure}

At zero temperature, for aspect ratios $\zeta$ larger than the critical ratio $\zeta_s=0.25$
and strong fields $H>J(1+4/L_y)$, the
interface is localized on the boundary, cf. \Fref{fig_pd_t0}. 
Although for $\zeta>\zeta_s$ no real solution of \Eref{eq_2} exists near $T_c$,
one can solve the equation for small
temperatures\footnote{Here we are in disagreement with Ref.~\cite{clufor} where no solution was found,
because of a mistake in the discriminant of \Eref{eq_2}.} 
and finds in the case $\zeta = 0.5$ the phase diagram shown in 
\Fref{fig_pd_zeta_0.5}. The zero temperature limit is consistent
with the energetic arguments in Ref.~\cite{clufor}, see also \Fref{fig_pd_t0}. The lower line starting 
at $T=0$, $H=1$ can also be detected by means of computer simulations, 
but as one can see in the right plot of \Fref{fig_pd_zeta_0.5}, for $T=1$ the 
dip between the two maxima in the boundary magnetization density $m_b=(1/L_y) \sum_{y=1}^{L_y} \sigma_{1,y}$
vanishes with increasing lattice sizes. 
Therefore, there is no signal for a first-order transition between these two regimes.
One can argue that there is no phase transition at all, because in the
infinite-volume limit only the ordered phase survives. The dashed line
in the left plot of \Fref{fig_pd_zeta_0.5} starting at $T=0$, $H=4\zeta=2$
is not visible in simulations,
because this second solution of \Eref{eq_2} would correspond to the boundary between
the bulk magnetized state and 
configurations with an interface propagating inside the bulk 
which, however, have a higher free energy then configurations with an interface 
localized near the boundary and hence are suppressed.
Therefore the phase diagram for $\zeta>0.25$ 
consists only of two phases, namely the ordered low-temperature phase
and the disordered high-temperature phase as in the pure 2D Ising model.
We hence conclude that for  $\zeta>0.25$ the inhomogeneous boundary 
magnetic field only leads to finite-size effects.

\begin{figure}[t]
\centerline{
\includegraphics[scale=1.15]{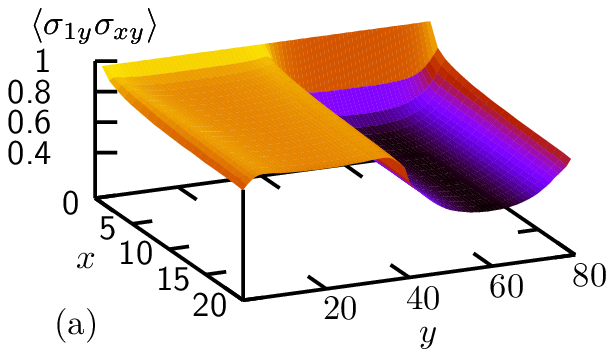}
\includegraphics[scale=1.15]{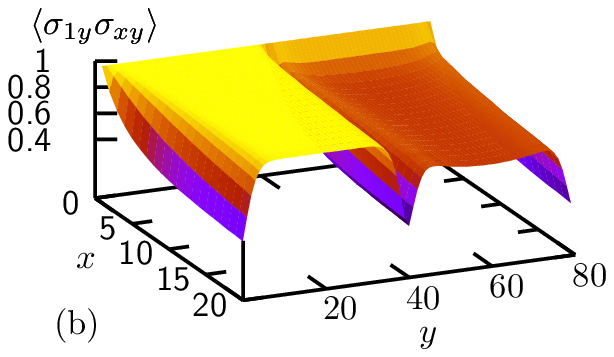}
}
\centerline{
\includegraphics[scale=1.15]{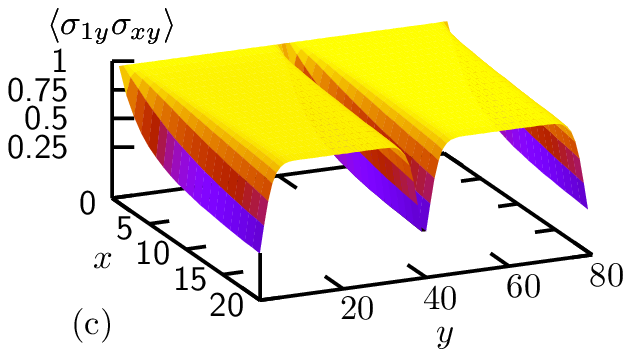}
}
\caption{Spin-spin correlation function $\langle \sigma_{1y}\sigma_{xy} \rangle$ for $\zeta=0.25$, $L_y=80$, and
$T=1$ measured for different values of the boundary magnetic field:
(a) $H=0.9$, 
(b) $H=0.925$, and  
(c) $H=0.975$. 
}
\label{fig_spin_spin_zeta_0.25}
\end{figure}

Finally, let us come to the special case of $\zeta=\zeta_s=0.25$. 
While the transition line disappears for $\zeta>1/4$ as the solutions 
of \Eref{eq_2} move to the complex plane, for $\zeta_s$ we still do 
see different peaks associated with the two phases and, therefore, 
a finite interface tension, cf. \Tref{tab1}.
The numerically estimated 
transition points for $H=0.5$ and $T=1.5$ also contained in \Tref{tab1}
are again seen to be in good agreement with \Eref{eq_2}. Furthermore, we 
analysed for $T=1.0$ the spin-spin correlation functions $\langle \sigma_{1y} \sigma_{xy}\rangle$
and $\langle \sigma_{x1} \sigma_{xy}\rangle$ in $x$- and $y$-direction, respectively.
Slightly below the transition from the ordered phase with $\langle m \rangle =m_0$
to the phase with an interface propagating inside the bulk and therefore 
$\langle m \rangle =0$, $\langle \sigma_{1y} \sigma_{xy}\rangle$ shows an asymmetric shape and a fast decay near the
boundary, indicating that the interface is localized near the boundary, cf.\ \Fref{fig_spin_spin_zeta_0.25} (a).
With increasing boundary magnetic field we cross the first-order line at $H_0\sim0.925$, where
the interface starts moving into the bulk and, therefore, the profile of $\langle \sigma_{1y} \sigma_{xy}\rangle$
becomes symmetric in $y$. Right at the transition line
where we have two coexisting phases we find both the fast decay  near the boundary as well as the
opening angle between the two almost symmetric halves, see \Fref{fig_spin_spin_zeta_0.25} (b).
When the boundary magnetic field is increased further, this mixed-phase effect vanishes and the 
interface in the bulk becomes stable. In this case we find a symmetric shape of the 
spin-spin correlation function, cf. \Fref{fig_spin_spin_zeta_0.25} (c),
similar to $\zeta=0.2$ in \Fref{fig_spin_spin} (b).

\section{Summary} \label{summary}

Our Monte Carlo data clearly confirm the theoretical considerations of  
Clusel and Fortin~\cite{clufor}
and extend their exact results by studying the cases $\zeta$ equal and
larger than the critical value $\zeta_s =1/4$. The observed finite-size
scaling behaviour fits nicely with their predictions for the infinite 
system, cf.\ our results in \Tref{tab1}. We also find that for a large 
aspect ratio some interesting finite-size effects can be observed, such as, for
example, a regime in the $H$--$T$ plane with two states separated by an
energy gap which vanishes in the infinite-volume limit. 
Furthermore, we studied the spin-spin correlation function for which
analytical results are not yet available. Since this observable turned out to
be quite sensitive to the interface location, it would be a challenging 
enterprise to pursue further analytical considerations in this direction.

\section*{Acknowledgments}
We gratefully acknowledge financial support from the Deutsche Forschungsgemeinschaft (DFG)
under Grant No. JA 483/23-1, 
the DFG Research Group FOR877,
and the EU RTN-Network `ENRAGE': {\em Random Geometry
and Random Matrices: From Quantum Gravity to Econophysics\/} under Grant
No.~MRTN-CT-2004-005616.

\newpage


\section*{References}
\begin{thebibliography}{99}

\bibitem{cahn} J.W. Cahn, J. Chem. Phys. {\bf 66}, 3667 (1977).
\bibitem{ebner} C. Ebner and W.F. Saam, Phys. Rev. Lett. {\bf 38}, 1486 (1977). 
\bibitem{nafi}  H. Nakanishi and M.E. Fisher, J. Chem. Phys. {\bf 78}, 3279 (1983).  
\bibitem{mccoy}
B.M. McCoy and T.T. Wu, {\em The Two-Dimensional Ising Model\/} (Harvard Univ. Press, Cambridge, Mass., 1973).
\bibitem{zamo} 
A.B. Zamolodchikov, Adv. Stud. Pure Math. {\bf 19}, 641 (1989);
Int. J. Mod. Phys. A {\bf 4}, 4235 (1989).
\bibitem{abra}
D.B. Abraham, Phys. Rev. Lett. {\bf 44}, 1165 (1980);
Phys. Rev. B {\bf 25}, 4922 (1982);
Phys. Rev. B {\bf 37}, 3835 (1988).
\bibitem{auyang}
H. Au-Yang and M.E. Fisher, Phys. Rev. B {\bf 11}, 3469 (1975).
\bibitem{clusel1} M. Clusel and J.-Y. Fortin, J. Phys. A {\bf 38}, 2849 (2005). 
\bibitem{clufor} M. Clusel and J.-Y. Fortin, J. Phys. A {\bf 39}, 995 (2006).
\bibitem{wj_nic}
W.~Janke, {\em Histograms and all that\/},
in: {\em Computer Simulations of Surfaces and Interfaces\/},
NATO Science Series, II.\ Mathematics, Physics and Chemistry -- Vol.~{\bf 114},
edited by B.~D\"unweg, D.P.~Landau, and A.I.~Milchev
(Kluwer, Dordrecht, 2003); pp.~137--157.
\bibitem{berg96}
B.A. Berg, J. Stat. Phys. {\bf 82}, 323 (1996).
\bibitem{1st_order}
W. Janke, {\em First-order phase transitions\/},
in: {\em Computer Simulations of Surfaces and Interfaces\/},
NATO Science Series, II. Mathematics, Physics and Chemistry -- Vol.~{\bf 114},
edited by B.~D\"unweg, D.P.~Landau, and A.I.~Milchev
(Kluwer, Dordrecht, 2003); pp.~111--135.


\end{thebibliography}
\end{document}